# TIME TRANSFER THROUGH OPTICAL FIBERS (TTTOF): PROGRESS ON CALIBRATED CLOCK COMPARISONS

Michael Rost[1], Miho Fujieda[2], Dirk Piester[1]

[1]Physikalisch-Technische Bundesanstalt (PTB)
Bundesallee 100, 38116 Braunschweig, Germany
E-mail: michael.rost@ptb.de

[2]National Institute of Information and Communications Technology (NICT)
Tokyo, Japan

*Abstract*

During the last years the transfer of frequency signals through optical fibers has shown ultra low instabilities in various configurations. The outstanding experimental results of such point-to-point connections is motivation to develop a means to extend the frequency transfer to accurate time transfer. We aim at the synchronization of clocks located at different places of the PTB campus with an over-all uncertainty of less than 100 ps. Such an installation can be used as a part of the infrastructure connecting local time scales with the ground station setup during forthcoming ACES experiments and the local two-way satellite time and frequency transfer (TWSTFT) installations.

This paper reports on the progress on time transfer through optical fibers (TTTOF) similar to the well known and long established TWSTFT scheme: A 10 MHz signal is transferred through an optical fiber connection to a remote site for realization of a time scale using a 10 MHz to 1PPS divider. For time transfer the 1PPS output is initially synchronized and then monitored by TWSTFT equipment connected by a second optical fiber to observe instabilities and uncertainties. Depending on further needs in future the transferred signal can be monitored only (for software correction) or controlled in real time by adding adequate phase shifters. We discuss procedures for a proper calibration of such TTTOF links and show results of experiments using fiber lengths up to two kilometers which prove that the proposed method is well suitable for the envisaged purpose.

## 1. INTRODUCTION

Transfer of frequency signals on distances up to more than 100 km have been examined and demonstrated recently [1]. Also very promising results have been reported for frequency transfer using standard 10 MHz signals [2] and phase transfer of a 1.5 GHz signal for radio astronomy applications [3]. These applications do not include the calibrated transfer of time signals to synchronize a remote clock to a local one.

As an extension of a previous study [4], we report here on the progress of a means for accurate time transfer using optical fibers. We aim at the synchronization of clocks located at different places on the institute campus of the Physikalisch-Technische Bundesanstalt (PTB). So, our target transmission length is below 1 km. Such an installation shall become part of the infrastructure connecting the ground station setups during the forthcoming ACES [5] experiment with the local installations at the time laboratory at the PTB. Additionally, the PTB TWSTFT ground stations, which are currently spread over the PTB campus, will be moved to a common location at the same new site. See Figure 1 and [6] for details. From late 2010 onwards the TWSTFT stations will be installed on top of a high building, where free sight to all directions is available. Especially for the ACES project it is necessary to establish a time scale in an extremely well known relation to UTC(PTB) in a "satellite time laboratory" next to the time transfer equipment for calibration and monitoring issues. The required uncertainty is 100 ps or less.

To cancel long-term fiber length variations the application of the two-way mode for exchanging optical signals has been demonstrated [7]. Such an approach was initially proposed in the framework of network synchronization [8]. Our time transfer through optical fibers (TTTOF) method is similar to the well known two-way satellite time and frequency transfer (TWSTFT) scheme which already has proven its feasibility for calibrated and operational satellite time transfer with the lowest uncertainty [9]. In both cases, code-domain-multiple-access (CDMA) signals are used.

## 2. CONCEPT AND SETUP

The basic concept for the installations in the remote "satellite time laboratory" includes the generation and distribution of a reference frequency (10 MHz) and time scale (1PPS) at the remote site. Both kinds of signals should be related to the reference clock at PTB's local time laboratory. In Figure 2 the basic setup is depicted. The reference frequency modulates an optical carrier signal by an electro-optical converter (E/O) and is transferred to the remote setup through a single mode optical fiber. At the remote site, it is transformed and distributed by an opto-electrical converter (O/E) and frequency distribution amplifier (FDA), respectively. We use two types of equipment: Ortel 10382S Fiberoptic Transmitter, 10481S Fiberoptic Receiver (manufacturer 1, MF1) and Linear Photonics Time Link Transmitters (TLTH1 SF) and receivers (TLRH1 SF) (manufacturer 2, MF2) for this purpose. The laser wavelength is nominally 1310 nm and 1550 nm, respectively. One output of the FDA feeds a divider (DIV) to provide 1PPS signals and the connected pulse distribution amplifier (PDA). One 1PPS output of the PDA is defined as the remote reference time scale TA(2). Initially, TA(2) has to be synchronized to TA(1) by a portable clock or by other suitable means. In principle, the phase difference between TA(2) and TA(1) is fixed, but affected by the delay instabilities on the transfer path. The delay change due to temperature variation in a spooled 1 km optical fiber, e.g., was reported to be about 30 ps/K/km [10]. The instability introduced by the one-way frequency transfer over an optical fiber of about only 1 km, however, is assumed to be small enough for our target, so that we omitted for now any phase correction systems in order to ensure a simple and reliable operational setup.

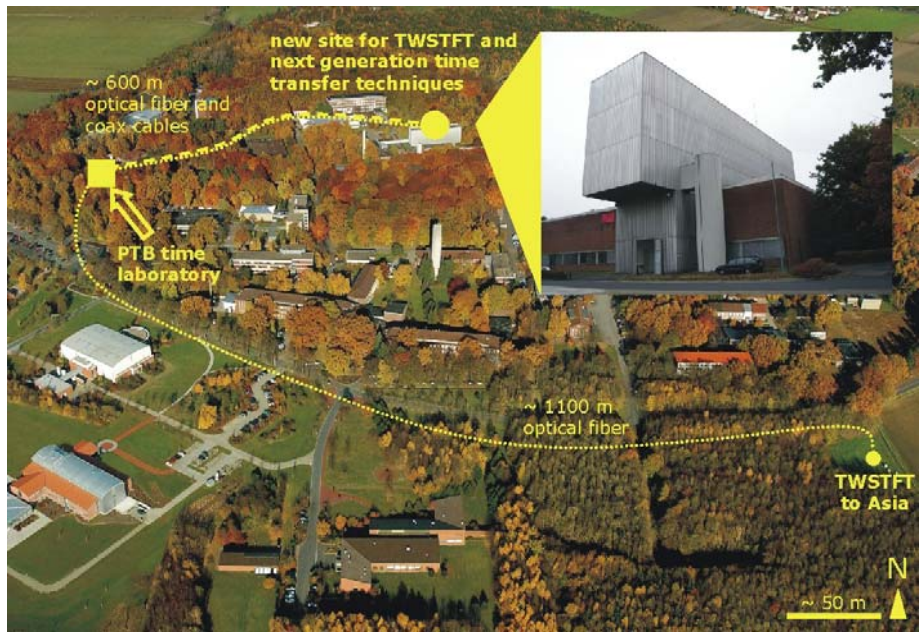

Fig. 1: Aerial view of the Braunschweig PTB campus, current optical fiber connection to the ground station for TWSTFT to Asia, and future installations to connect the new site for TWSTFT and next generation time transfer techniques, e.g. the ACES ground terminal, with the PTB time laboratory.

We measure the difference between TA(2) and TA(1) using TWSTFT modems (type TimeTech SATRE) which are widely used in many time laboratories. In operation the modems at each site exchange binary phase-shift keyed (BPSK) signals. Each modem transmits a 70 MHz signal modulated with a 20 MCh/s BPSK sequence phase coherent to its reference TA(*i*). In both modems the phase offset between the signal sent from the counterpart modem and the local reference is measured by the cross correlation method. From these measurements the difference TA(1) – TA(2) can be computed after suitable calibration of the equipment internal delays as described below. The signals from the modems are transformed by using MF1 E/Os and O/Es as mentioned above, but with laser wavelengths of nominally 1550 nm. Additionally we used standard telecommunication circulators and isolators. For clarity the isolators and also additional electrical amplifiers or attenuators are not displayed in the figures throughout this paper.

We use additional experimental equipment for the characterization of the time and frequency transfer. In Section 3 the instability of both the one-way frequency transfer as well as the two-way time transfer are characterized by using two different phase comparators type VREMYA-CH VCH-312 and VCH-314 and for tests a time interval counter of type

SRS SR620. The measurement result of the two-way time transfer should be independent of the length of the optical fiber between the two circulators. As discussed in Section 4, we are using different fiber lengths, i.e. a test loop using a short fiber (2 to 15 m) and a loop of overall 2.2 km length on the PTB campus (see Figure 1) to verify this assumption.

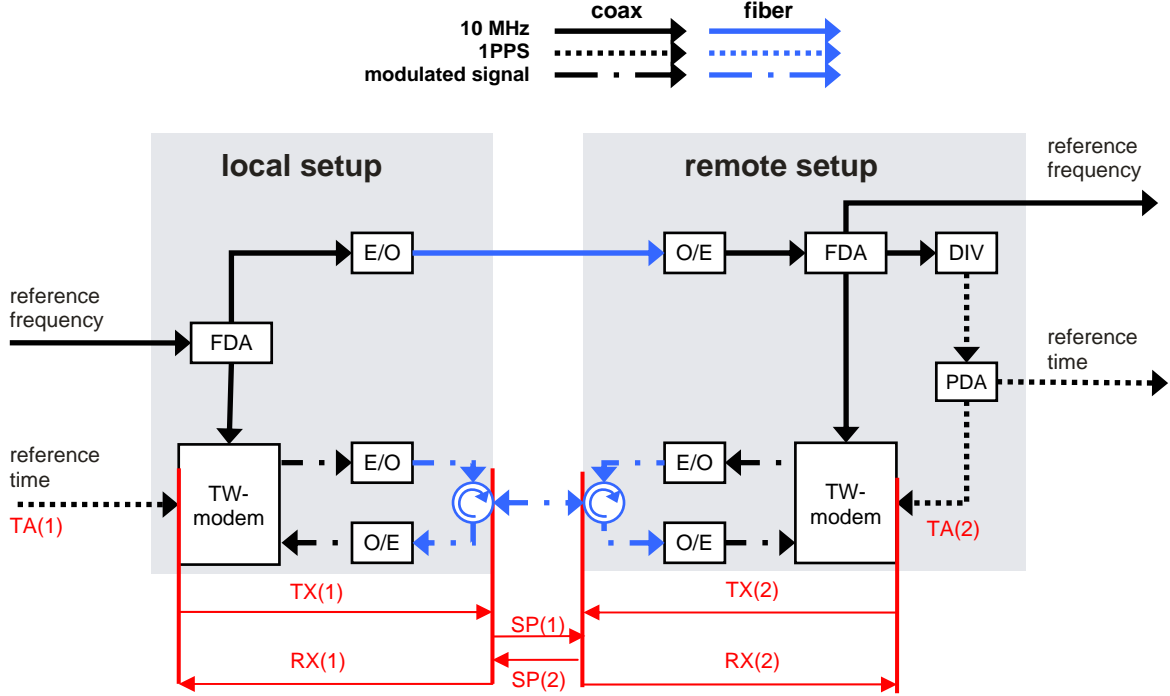

Figure 2: The concept: A remote time scale is generated using a one-way frequency transfer. The time scale is monitored by a two-way measurement system employing modems usually used for TWSTFT. Red markings define the delays used in the text.

We want to compare the remote time scale TA(2) with the local one TA(1). If possible the nomenclature of existing literature [9, 11] is applied in the following and we only deviate if necessary. We start with the two-way measurements answers as provided by each of the TWSTFT modems

$$TW(1) = TA(1) - TA(2) + TX(2) + SP(2) + RX(1), \qquad (1)$$
$$TW(2) = TA(2) - TA(1) + TX(1) + SP(1) + RX(2). \qquad (2)$$

TX(i) and RX(i) represent the complete internal transmission and receive delay of setup i. SP(1) is the transmission path delay from the local setup to the remote setup. SP(2) is the delay in opposite direction through the same fiber. As mentioned above, these measurements are time of arrival measurements of a signal from the remote site with respect to the local reference, and vice versa. For a single optical fiber connection between the two sites we assume SP(1) = SP(2), and thus

$$TA(1) - TA(2) = \tfrac{1}{2}\,[TW(1) - TW(2)] + \tfrac{1}{2}\,[[TX(1) - RX(1)] - [TX(2) - RX(2)]] \qquad (3)$$

is valid. Introducing the delay difference between TX and RX path of one setup, DLD(i) = TX(i) – RX(i), (3) simplifies to

$$TA(1) - TA(2) = \tfrac{1}{2}\,[TW(1) - TW(2)] + \tfrac{1}{2}\,[DLD(1) - DLD(2)]. \qquad (4)$$

For a accurate time transfer we need to determine the second term in (4). To calibrate the setups we use a common clock configuration and adjust TA(1) = TA(2), which leads to

$$0 = \tfrac{1}{2}\,[TW(1) - TW(2)] + \tfrac{1}{2}\,[DLD(1) - DLD(2)]. \qquad (5)$$

The first term is called common clock difference CCD(1,2) = ½ [TW(1) – TW(2)]. It is measured to determine the relative delay difference between both systems

$$CCD(1,2) = -½ [DLD(1) - DLD(2)]. \quad (6)$$

This relative delay difference is exactly the calibration value CALR(1,2) for the link between the two setups

$$TA(1) - TA(2) = ½ [TW(1) - TW(2)] + CALR(1,2). \quad (7)$$

A detailed absolute determination of the single TX(i) and RX(i) delays is not necessary.

## 3. FREQUENCY INSTABILITY

The instability of the frequency transfer system was characterized using the laboratory setup as depicted in Figure 3. We used an optical fiber link of 2 km buried in the PTB campus for the measurements (see Figure 1). A reference frequency from PTBs hydrogen maser H5 is transferred from the "local" frequency distribution amplifier (FDA) (left hand in Figure 3) to the "remote" one (right hand in Figure 3) via the 2 km optical fiber test loop or a short test fiber. Different optical equipment have been tested for this link (MF1 and MF2). A phase comparator is connected directly to both FDAs and measures the phase difference between the two frequency signals. The two-way modems need DIVs after both FDAs to feed the modems with 1PPS signals. Here the time difference between the two 1PPS signals is measured by the modems. We used a second 2 km fiber loop of the same buried cable as above for the two-way measurement system. A test length of 2 km allows a characterization of the future setup in a realistic scenario.

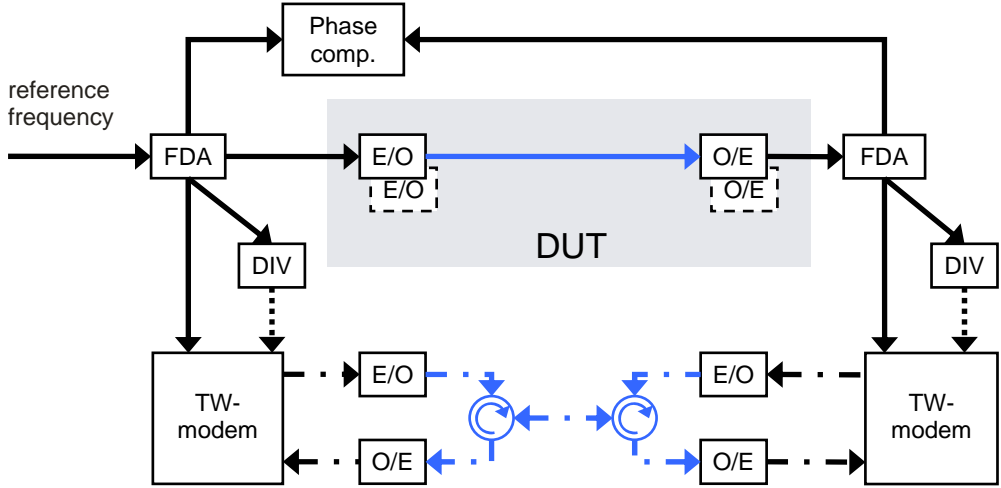

Figure 3: Setup to measure the instability of the one-way fiber frequency transfer
as well as the instability of the two-way monitor measurements by means of a phase comparator.

Results of the measurements are depicted in Figure 4. The figure shows instability graphs from the one-way 10 MHz frequency transfer links measured with a phase comparator as well as results from the monitoring TWSTFT system. MF1 measurements (two-way and phase comparator) show similar results after an averaging time of 100 s. The two-way measurements exhibit significant frequency instability of $6·10^{-12}$ at an averaging time of 1 s. This is in good agreement with the noise level one would expect for this type of modem operated at 20 MCh/s BPSK [12]. Note that TW measurements have been performed only with the MF1 components and may be improved when using MF2 components at least averaging times longer than 100 s.

The phase comparator measurements have significantly lower frequency instability at 1 s averaging time. The results for the MF2 components show a better stability at some averaging times almost by a factor of 10.

As a summary, one-way frequency transfer via the optical fiber shows an instability below $10^{-15}$ at $10^4$ s averaging using MF1 components and an instability at the $10^{-16}$ level with MF2 components. Tests for the two-way measurements using the MF2 components will be performed in the near future. We expect the monitoring system to go down to instabilities below the red graph in Figure 4 (MF1) at averaging times of more than 100 s to values similar to the instabilities of frequency transfer using the MF2 components. The short-term stability of the TW measurement is limited by the chip rate of the modem hardware of 20 Mch/s. If the instability plot follows the expected slope, it will approach the instability of MF2 components at averaging times about 1000 s.

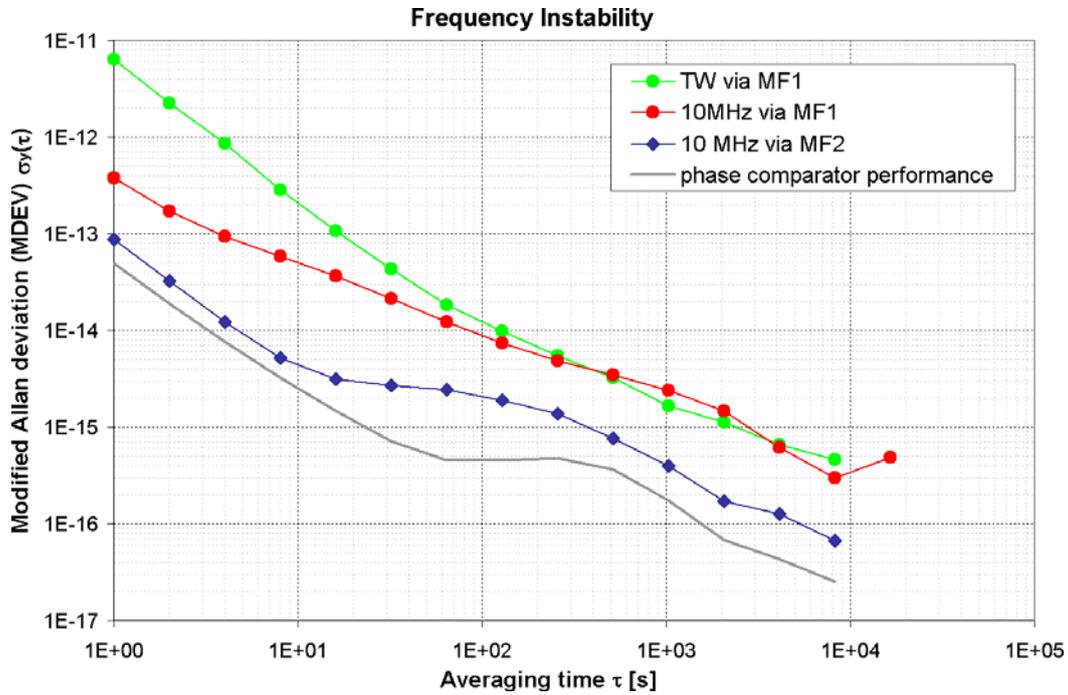

Figure 4: Fractional frequency instability of the two-way measurements (green circles, via MF1 components only) [4], phase comparator measurements of MF1 components (red cycles) and MF2 components (blue diamonds). The measured phase comparator performance is depicted as a gray line.

## 4. TIME ACCURACY

A prerequisite for accurate comparisons of remote time scales is the possibility for delay calibration of the whole system. Because the optical fiber length of the final setup is unknown, a calibration test is needed to ensure the independence of the setup from the length of the used fiber. For this purpose we connected the modems to reference frequency and 1PPS from FDAs and DIVs as illustrated in Figure 5, and changed the length of the fibers. We chose different fiber lengths (the indoor fiber is 2m, the outdoor loop is 2 km long). The attenuation of the two optical fibers was chosen to be at the same level by inserting a variable optical attenuator into the fiber of common signal path (SP(1), SP(2)) to minimize the impact of receive power dependent delay variations in the modems. The optical power was kept constant within ±0.1 dB. All accuracy tests have been performed with MF1 components so far, an accuracy test with MF2 components is planned for near future.

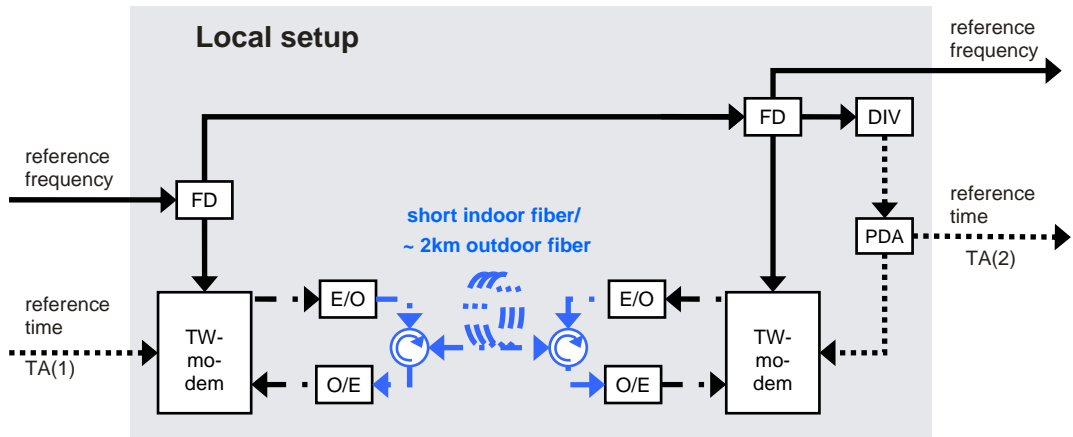

Figure 5: Calibration setup for testing the independence of the time transfer results from the length of the optical fiber.

The results of CCD measurements (6) when switching over between long and short fibers is depicted in Figure 6. The sequence comprises 8 switches between the long and the short fiber. The error bars in Figure 6 represent the standard deviation of single measurements. The higher standard deviation of the measurements with the short fiber may be due to instabilities caused by interference. Nevertheless, the standard deviation of the 9 results around the mean is only 6 ps. Variations at the beginning of the sequence might be result of temperature variations which have to be studied in a long term experiment. Variations of less than 40 ps (including error bars) under different experimental conditions is a promising result to comply with the aim of enabling time transfer with an uncertainty well below 100 ps.

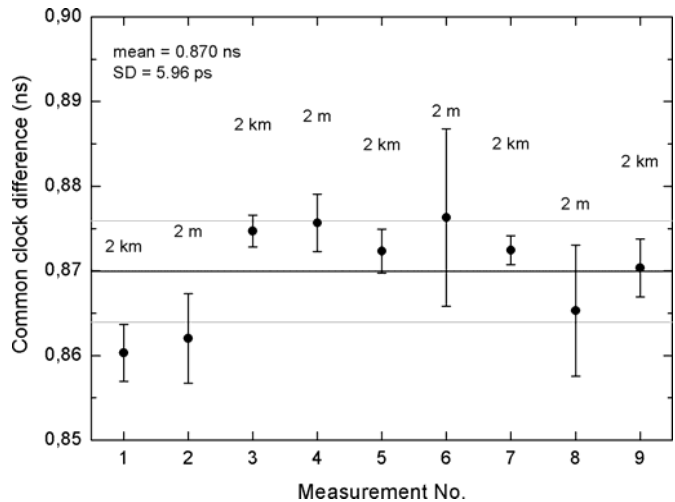

Figure 6: Display of the sequences of measurements with short indoor fiber and long outdoor fiber loop.

Additionally we performed a switch off-and-on sequence of all involved equipment to simulate the anticipated transfer of the setup to the remote location. The electrical power of all involved devices was switched off for about one hour and switched on again to simulate breaks of operation during installation at new "Satellite Laboratory" and due to maintenance of PTBs internal fiber network. This was repeated 3 times. The results of the measurements are displayed in Figure 7: A good reproducibility with a standard deviation of ± 6.2 ps was observed.

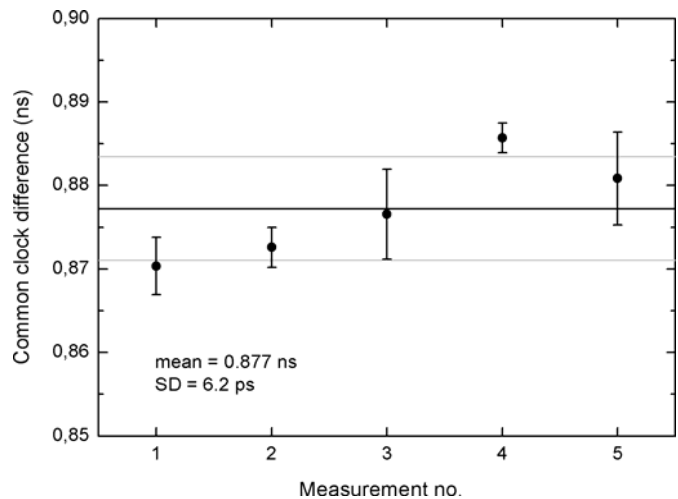

Figure 7: Display of the sequence of off-and-on measurements results.

## 5. SUMMARY

We have developed a method to monitor a remote time scale by means of a two-way optical fiber time transfer system. This method is combined with a one-way fiber transfer of the reference frequency to generate the time scale at the remote site with sufficient instability. Because our goal is to establish an institute campus solution with distances of about 1 km we have not introduced any phase variation compensation system but rely on monitoring of the fiber related instabilities. in a previous study [4] we have shown that the one-way frequency transfer via an optical fiber length of 2 km is possible at the $10^{-15}$ level after $10^4$ s averaging. The two-way monitoring system shows an instability well below $10^{-15}$ after $10^4$ s averaging. Here we have demonstrated that different commercially available hardware components can improve the one-way frequency transfer by a factor 5. For accurate time transfer we tested in detail the independence of the time transfer results from the transmission path. A variation of less than 40 ps under different experimental conditions has been reconfirmed in any case. The crucial point of receive power dependence of the modems' internal delays which has been taken into account by regulating the optical power on the two-way path to keep the receive power of the modems constant within ± 0.1 dB.

Future investigations will address the long term validation of the time transfer accuracy, as well as other concepts for the two-way monitor scheme, e.g. different wavelengths of the lasers in the two-way monitor system [7]. Also phase compensation systems, like a phase shifter or fiber heater [13], derived from the real time solutions delivered by the SATRE modems' data outputs could help to improve the stability of the one-way frequency transfer.

## ACKNOWLEDGEMENT

The authors thank Wolfgang Schäfer (TimeTech GmbH) for the loan of one SATRE modem and helpful discussions.

## DISCLAIMER
The Physikalisch-Technische Bundesanstalt as a matter of policy does not endorse any commercial product. The mentioning of brands and individual models seems justified here, because all information provided is based on publicly available material or data taken at PTB and it will help the reader to make comparisons with own observations.